\documentclass[twocolumn,showpacs,preprintnumbers,amsmath,amssymb,floatfix]{revtex4}

\usepackage{graphicx}
\usepackage{dcolumn}
\usepackage{bm}

\newcommand{\beq}{\begin{equation}}
\newcommand{\eeq}{\end{equation}}
\newcommand{\bey}{\begin{eqnarray}}
\newcommand{\eey}{\end{eqnarray}}

\begin{document}

\title{Singularity-free solutions for anisotropic charged fluids with Chaplygin equation of state}

\author{Farook Rahaman}
\email{farook\_rahaman@yahoo.com} \affiliation{Department of
Mathematics, Jadavpur University, Kolkata 700 032, West Bengal,
India}

\author{Saibal Ray}
\email{saibal@iucaa.ernet.in} \affiliation{Department of Physics,
Government College of Engineering \& Ceramic Thechnology, Kolkata
700 010, West Bengal, India}

\author{Abdul Kayum Jafry}
\email{akjafry@yahoo.com} \affiliation{Department of Physics,
Shibpur Dinobundhoo Institution, Howrah - 700 102, West Bengal,
India}

\author{Kausik Chakraborty}
\email{kchakraborty28@yahoo.com}
\affiliation{Department of
Physics, Government Training College, Hooghly 712103, West Bengal,
India}

\date{today}

\begin{abstract}
We extend the Krori-Barua analysis of the static, spherically
symmetric, Einstein-Maxwell field equations and consider charged
fluid sources with anisotropic stresses. The inclusion of a new
variable (tangential pressure) allows the use of a non-linear,
Chaplygin-type equation of state with coefficients fixed by the
matching conditions at the boundary of the source. Some physical
features are briefly discussed.
\end{abstract}


\maketitle

\section{Introduction}
A major issue of the static, spherically symmetric Einstein field
equations in GR is to find out interior solutions which are free
of singularity. It has shown that an uncharged incompressible
fluid sphere of mass $m$ cannot be held in equilibrium below
certain radius $a = \frac{9m}{4}$ and even demands a larger value
for $a$ related to physically reasonable equation of state (EOS)
\cite{Buchdhal1959}. Regarding stability of the model Stettner
\cite{Stettner1973} argued that a fluid sphere of uniform density
with a net surface charge is more stable than without charge.
Therefore, a general mechanism have been adopted by the
relativists to overcome this singularity due to gravitational
collapsing of a static, spherically symmetric fluid sphere is to
include charge to the neutral system. It is observed that in the
presence of charge either gravitational attraction is counter
balanced by the electrical repulsion in addition to the pressure
gradient \cite{Sharma2001} or inhibits the growth of space-time
curvature which has a great role to avoid singularities
\cite{Felice1995}.  According to Ivanov \cite{Ivanov2002} the
presence of the charge function serves as a safety valve, which
absorbs much of the fine tuning, necessary in the uncharged case.
However, in connection to this we would like to mention here a
special kind of mechanism to avert singularity as used by Trautman
\cite{Trautman19} under Einstein-Cartan theory where physical
entity spin-torsion supposed to act as an agent of repulsive
effect.

A large amount of works on charged fluid spheres are available in
the literature  (an exhaustive discussion with various
classification schemes regarding sources for the
Reissner-Nordstr\"{o}m (RN) space-time can be obtained in Ref.
\cite{Ivanov2002}). However, in connection to singularity we
would like to mention here that Efinger \cite{Efinger1965}, Kyle
and Martin \cite{Kyle1967} and Wilson \cite{Wilson1967} have found
relativistic internal solutions for static charged spheres, but
none of these solutions is absolutely free from singularities. On
the other hand, spheres of charged dust have been investigated by
Bonnor \cite{Bonnor1965}, Bonnor and Wickramasuriya
\cite{Bonnor1975} and Raychaudhuri \cite{Raychaudhuri1975}. Among
all these investigations it is observed that in Efinger's solution
the metric has a singularity at the origin ($r = 0$) whereas the
solutions due to Kyle and Martin \cite{Kyle1967} and Wilson
\cite{Wilson1967} do not have any interior singularities. However,
it is argued by Junevicus \cite{Junevicus1976} that in both the
above cases the metrics may have singularities at points other
than the origin so that restrictions have to be imposed on the
sphere to avoid them. According to him the fluid sphere solutions
of Kyle and Martin \cite{Kyle1967}, Wilson \cite{Wilson1967},
Kramer and Neugebauer \cite{Kramer1971} and Krori and Barua
\cite{Krori1975} are of special interest since, with the
imposition of suitable conditions, they are completely free of
metric singularities and satisfy physical considerations (for the
discussion and analysis of stability vide Refs. \cite{Ray2007a}
and \cite{Ray2007b} respectively).

We would like to note here, specially, the works of Krori and
Barua (KB) \cite{Krori1975} and Junevicus \cite{Junevicus1976}
which are the basis of our present investigation. Krori and Barua
(KB) \cite{Krori1975} constructed static, spherically symmetric
solutions of the Einstein-Maxwell equations based on a particular
choice of the metric components $g_{00}$ and $g_{11}$ in curvature
coordinates. Assuming that the source is a charged fluid with
isotropic stresses, the three independent Einstein equations were
reduced to linear algebraic equations for the energy density
$\rho(r)$, pressure $p(r)$ and the square of the electric field,
$E(r)^2$. In this approach the independent Maxwell equation is
used to obtain the charge density from the pre-determined form of
$E(r)$. A special feature of KB \cite{Krori1975} solutions is that
they are singularity free. A thorough analysis of this
singularity-free KB \cite{Krori1975} solution has been done by
Junevicus \cite{Junevicus1976}. The main aspect of his
investigation is to fix up the constants involved in the KB
\cite{Krori1975} metric in terms of the physical constants of
mass, charge and radius of the source. He also investigated the
conditions for physical relevance leading to a functional
dependence of the ratio of mass-to-radius on the ratio of
charge-to-mass and also to upper and lower limits on these ratios.
In his recent work on static charged perfect fluid spheres in
general relativity Ivanov \cite{Ivanov2002} has also observed that
the solution of KB \cite{Krori1975} which are fixed by the
junction conditions is non-singular and the positivity conditions
are satisfied.

In connection to the above discussion on KB solution
\cite{Krori1975} it is to be mentioned regarding the very recent
work on these solution by Varela et al. \cite{Varela2010}. The
work deals with self-gravitating, charged and anisotropic fluids
to solve the Einstein-Maxwell equations. In order to discuss
analytical solutions they \cite{Krori1975} extend KB method
\cite{Krori1975} to include pressure anisotropy and linear or
non-linear equations of state. The obtained solutions satisfy the
energy conditions of general relativity and have the following
features: (1) spheres with vanishing net charge contain fluid
elements with unbounded proper charge density located at the
fluid-vacuum interface; (2) inward-directed fluid forces caused by
pressure anisotropy may allow equilibrium configurations with
larger net charges and electric field intensities than those found
in studies of charged isotropic fluids; (3) links of these results
with charged strange quark stars as well as models of dark matter
including massive charged particles are possible, and (4) the Van
der Waals equation of state leading to matter densities
constrained by cubic polynomial equations is considered.

Our present investigation of static, spherically symmetric charged
fluid sphere distribution is in continuation of the above work of
anisotropic fluid source \cite{Varela2010}. This means that the
radial and tangential pressures are, in general, unequal so that
the simplest relation between them may be assumed as $p_t = n p_r,
\quad (n\neq 1)$ \cite{Ray2008}. However, there are several other
forms of anisotropic relationship between pressures can be noted
in the literature, e.g., in connection to electromagnetic mass
model Herrera and Varela \cite{Herrera1994} introduced a condition
of anisotropy in the form $p_t - p_r = g q^2 r^2$ where $g$ is a
constant having non-zero value whereas Barreto et al.
\cite{Barreto2006} define the degree of local anisotropy induced
by charge as $p_t - p_r = \frac{E^2}{4\pi}$, where $E$ is the
local electric field intensity, to consider self-similar and
charged radiating fluid spheres as anisotropic sources.

Recently, scientists show great interest on Chaplygin gas EOS in
order to explain accelerating phase of the present Universe as
well as to unify the dark energy and dark matter. As Chaplygin gas
EOS is a specific form of polytropic EOS so it looks promising to
describe dark energy spherically symmetric charged objects,
generally termed as {\it dark stars} in the literature
\cite{Bertolami2005,Cattoen2005,Lobo2006,Chan2008a}. As a possible
mechanism of formation it is argued by several investigators that
the first stars to form in the Universe, at redshifts $z \sim 10 -
50$, may be powered by dark matter annihilation for a significant
period of time rather than nuclear fusion
\cite{Spolyar2008,Freese2008,Scott2008,Schleicher2009}. On the
other hand, it is believed that dark energy exerts a repulsive
force on its surrounding and this repulsive force, likewise
electric charge, may prevent the star from gravitational collapse.
Therefore, people have speculated that a massive star does not
simply collapse to form a black hole, instead to the formation of
dark energy star with a final configuration without neither
singularities nor horizons
\cite{Ferreira1997,Ma1999,Mazur2002,Mota2004}.

However, among the above mentioned dark star models we are
specially interested to the work of Bertolami and P{\'a}ramos
\cite{Bertolami2005} where they, like us, have used the
generalized Chaplygin gas (GCG) EOS in special reference with
anisotropic pressure though our motivation and approach to solve
the spherically symmetric gas model is quite different from them.
The scheme of the present investigation is therefore as follows:
we write down the four independent Einstein-Maxwell equations. By
allowing the radial ($p_r$) and tangential ($p_t$) pressures to be
different, we have found out the six variables $p_r$, $p_t$,
$\rho$, $\epsilon=E(r)^2$, $\lambda$, $\nu$ where the other
parameters are, respectively, matter-energy density, electric
field intensity and metric potentials (Sec. II).  Ivanov
\cite{Ivanov2002} has explained the usual difficulties that
generally arise when we combine equations of state (even a linear
one) with the field equations. Interestingly, we are here dealing
with a non-linear EOS and are able to find solutions using an
algebraic method (we do not solve differential equations). Adding
the non-linear Chaplygin gas EOS $p_r=H\rho-\frac{K}{\rho}$ (where
$H$ and $K$ are two arbitrary constants) and using the KB ansatz
for $\lambda$ and $\nu$ we get four algebraic equations for
$\rho$, $p_r$, $p_t$, $\epsilon=E(r)^2$. Using the independent
Maxwell equation we determine the charge density $\sigma$ from
$\epsilon$ (Sec. III). We present and discuss the necessary
matching of the solutions and the related boundary conditions in
the Sec. IV which allow us to find out the expressions for $H$ and
$K$ with their physical features through the graphical plots.
Also, energy conditions have been discussed in detailed (Sec. V).
Some concluding remarks are made in Sec. VI.

\section{Basic Equations}
The KB \cite{Krori1975} metric is given by
\begin{equation}
ds^{2}=-e^{\nu(r)}dt^2+e^{\lambda(r)}dr^2+r^2(d\theta^2+\sin^2\theta
d\phi^2) \label{eq:kbm}
\end{equation}
where $\lambda(r)=Ar^2$ and $\nu(r)=Br^2+C$ with arbitrary
constants $A$, $B$ and $C$.

The most general energy momentum tensor compatible with
spherically symmetry is
\begin{equation}
T_\nu^\mu=  ( \rho + p_r)u^{\mu}u_{\nu} - p_r g^{\mu}_{\nu}+
            (p_t -p_r )\eta^{\mu}\eta_{\nu} \label{eq:emten}
\end{equation}
with $$ u^{\mu}u_{\mu} = - \eta^{\mu}\eta_{\mu} = 1 $$.

The Einstein-Maxwell equations are
\begin{equation}e^{-\lambda}
\left[\frac{\lambda^\prime}{r} - \frac{1}{r^2}
\right]+\frac{1}{r^2}= 8\pi \rho + E^2, \label{eq:lam}
\end{equation}
\begin{equation}e^{-\lambda}
\left[\frac{1}{r^2}+\frac{\nu^\prime}{r}\right]-\frac{1}{r^2}=
8\pi p_r - E^2, \label{eq:nu}
\end{equation}
\begin{equation}\frac{1}{2} e^{-\lambda}
\left[\frac{1}{2}(\nu^\prime)^2+ \nu^{\prime\prime}
-\frac{1}{2}\lambda^\prime\nu^\prime + \frac{1}{r}({\nu^\prime-
\lambda^\prime})\right] =8\pi p_t + E^2 \label{eq:tan}
\end{equation}
and
\begin{equation}
(r^2E)^\prime = 4\pi r^2 \sigma e^{\frac{\lambda}{2}}.
\label{eq:elec1}
\end{equation}
Equation (6) can equivalently be expressed in the form
\begin{equation}
E(r) = \frac{1}{r^2}\int_0^r 4\pi r^2 \sigma
e^{\frac{\lambda}{2}}dr = \frac{q(r)}{r^2} \label{eq:elec2}
\end{equation}
where $q(r)$ is the total charge of the sphere under
consideration.\\

\section{Solutions}
Now, we consider KB ansatz:
\begin{equation}
\lambda(r)=Ar^2, ~ \nu(r)= Br^2+C \label{eq:kba}
\end{equation}
where, as mentioned earlier, $A$, $B$ and $C$ are some arbitrary
constants. It is of interest to note that these constants were
determined by Junevicus \cite{Junevicus1976} in terms of the
physical quantities mass, charge and radius of the source.

Along with the above ansatz let us also use generalized Chaplygin
gas EOS for the charged fluid as \cite{Benaoum2002}
\begin{equation}
p_r = H\rho -\frac{K}{\rho} \label{eq:rad}
\end{equation}
where $H$ and $K$ are two positive constants.

Equations (\ref{eq:lam}) and (\ref{eq:nu}) implies
\begin{equation}
(\rho+p_r) \equiv f(r) = \frac{(A+B)}{4\pi} e^{-Ar^2}.
\label{eq:eos}
\end{equation}

From equations (\ref{eq:rad}), (\ref{eq:kba}) and (\ref{eq:eos}),
we get the following solution set:

\begin{equation}
\rho = \left(\frac{f +\sqrt{f^2 + 2hK}}{h}\right), \label{eq:rho}
\end{equation}

\begin{equation}
p_r = f - \left(\frac{f +\sqrt{f^2 + 2hK}}{h}\right),
\label{eq:p_r}
\end{equation}

\begin{widetext}
\begin{eqnarray}
p_t =\frac{1}{8\pi}\left[2e^{-Ar^2}(B-A)(2+Br^2)
-\frac{1}{r^2}\left(1- e^{-Ar^2}\right) + 8\pi \left(\frac{f
+\sqrt{f^2 + 2hK}}{h}\right)\right],\label{eq:p_t}
\end{eqnarray}
\end{widetext}

\begin{widetext}
\begin{equation}
E^2 = 2A e^{-Ar^2}+\frac{1}{r^2}(1- e^{-Ar^2})-8\pi \left(\frac{f
+\sqrt{f^2 + 2hK}}{h}\right), \label{eq:E^2}
\end{equation}
\end{widetext}

\begin{widetext}
\begin{equation}
q^2 = 2A r^4e^{-Ar^2}+r^2(1- e^{-Ar^2})-8\pi r^4 \left[\frac{f
+\sqrt{f^2 + 2hK}}{h}\right] \label{eq:q^2}
\end{equation}
\end{widetext}

where $h=2(1+H)$. We observe that for finite values of the
physical parameters $h \neq 0$, so that $H \neq -1$.

\section{Boundary conditions}
The RN metric \cite{Reissner1916,Nordstrom1918} is given by
\begin{widetext}
\begin{equation}
ds^{2}=-\left(1 - \frac{2m}{r} + \frac {Q^2}{r^2}\right)dt^2 +
\left(1 - \frac{2m}{r} + \frac {Q^2}{r^2}\right)^{-1}dr^2 +
r^2(d\theta^2+\sin^2\theta d\phi^2). \label{eq:rnm}
\end{equation}
\end{widetext}
To match our interior metric with the above exterior one we impose
only the continuity of $g_{tt}$, $g_{rr}$ and $\frac{\partial
g_{tt}}{\partial r}$ across a surface, $S$, at $r= a$. This yield
the following equations:
\begin{equation}
1 - \frac{2m}{a} + \frac {Q^2}{a^2} = e^{Ba^2+C},
\end{equation}
\begin{equation}
1 - \frac{2m}{a} + \frac {Q^2}{a^2} = e^{Aa^2},
\end{equation}
\begin{equation}
\frac{m}{a^2} - \frac {Q^2}{a^3} = Bae^{Ba^2+C}.
\end{equation}
Therefore, from the above equations, one can easily get
\begin{equation}
A = - \frac{1}{a^2} \ln \left[ 1 - \frac{2m}{a} + \frac {Q^2}{a^2}
\right], \label{eq:A}
\end{equation}
\begin{equation}
B = \frac{1}{a^2} \left[\frac{m}{a} - \frac {Q^2}{a^2}\right]
\left[1 - \frac{2m}{a} + \frac {Q^2}{a^2} \right]^{-1},
\label{eq:B}
\end{equation}
\begin{equation}
C =  \ln \left[ 1 - \frac{2m}{a} + \frac {Q^2}{a^2} \right]-
\frac{ \frac{m}{a} - \frac {Q^2}{a^2}}{ \left[ 1 - \frac{2m}{a} +
\frac {Q^2}{a^2} \right]}. \label{eq:C}
\end{equation}

Note that no extra assumption on the value of $\epsilon(r)$ at
$r=a$ is required here and we therefore obtain
$\epsilon(a)=\frac{Q^2}{a^4}$. This result was expected as a
consequence of the matching conditions at $r=a$ (absence of thin
shell). On the other hand, the electric field at $r=0$ is zero.
Therefore, the energy conditions at $r=0$ involve only the central
value of density, as well as for the radial and tangential
pressures. As thin shell exists, so we could not use boundary
condition for $\epsilon$ at $a$ which means one can not get
$\epsilon(a) = \frac{Q^2}{a^4}$.

Let us now impose the boundary conditions
\begin{equation}
p_r(a) = 0, ~\epsilon(0)= E(0) = 0.
\end{equation}
We obtain two independent equations which are readily solved for
$H$ and $K$ as functions of the source parameters. We note that
$E(0) = 0$ implies
\begin{equation}
\rho_0=\frac{3A}{8\pi} = \left[\frac{Fe^{Aa^2}
+\sqrt{\left(Fe^{Aa^2}\right)^2 + 2hK}}{h}\right] \label{eq:rho01}
\end{equation}
and $p_{r}(a)= 0$ implies
\begin{equation}
F -\left[\frac{F+\sqrt{F^2 + 2hK}}{h}\right] =0 \label{eq:rad0}
\end{equation}
where $F= f(a)= \frac{(A+B)}{4\pi}e^{-Aa^2}$.

From the above two equations (\ref{eq:rho01}) and (\ref{eq:rad0}),
one could find the values of two unknowns $H$ and $K$ in terms of
$A$, $B$ and $a$, in other words, in terms of mass, charge and
radius of the spherically symmetric charged objects. Therefore,
through a simple mathematical exercise, we have the expressions
for the constants as follows

\begin{equation}
H = \frac{3A(B-A)}{[9A^2-4(A+B)^2 e^{-2Aa^2}]}  \label{eq:h}
\end{equation}

\begin{widetext}
\begin{equation}
K= \left(\frac{(A+B)
e^{-Aa^2}}{4\pi}\right)^2\left[2\left(1+\frac{3A(B-A)}{[9A^2-4(A+B)^2
e^{-2Aa^2}]} \right)^2 -1 \right]\label{eq:k}
\end{equation}
\end{widetext}

One can note that at r = 0, E(0) = 0, $\rho(0)  =
\frac{3A}{8\pi}$, $p_r(0)=\frac{1}{2}p_t(0) =\frac{2B-A}{8\pi}$.
Also, the curve profiles (Figs 1 - 3) for the parameters, $\rho$,
$p_r$ etc. indicate no singularity presents inside the star.

\begin{figure*}[ptbh]
\includegraphics[scale=.4]{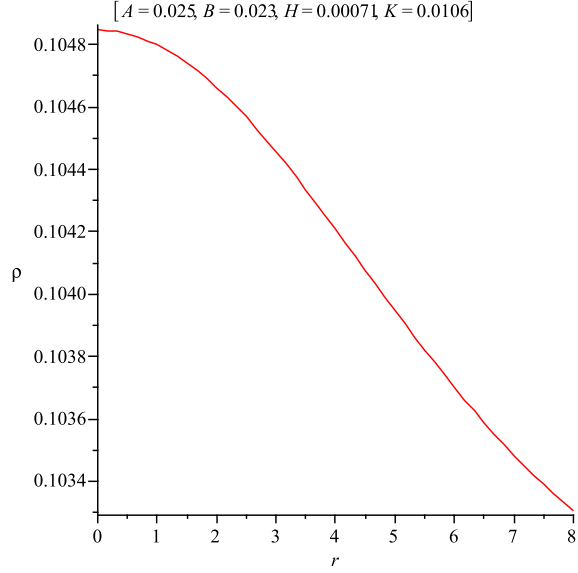}
\caption{The density parameter $\rho$ is shown against $r$.}
\label{Fig. 1}
\end{figure*}

\begin{figure*}[htbp]
\includegraphics[scale=.4]{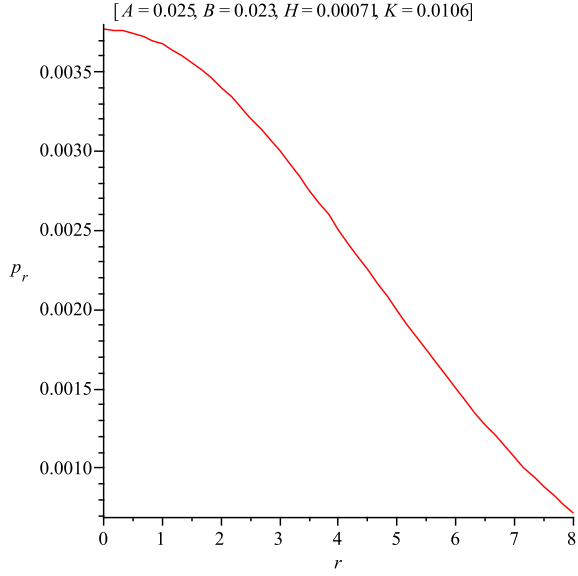}
\caption{The radial pressure $p_r$ is shown against $r$.}
\label{Fig. 2}
\end{figure*}

\begin{figure*}[htbp]
\includegraphics[scale=.4]{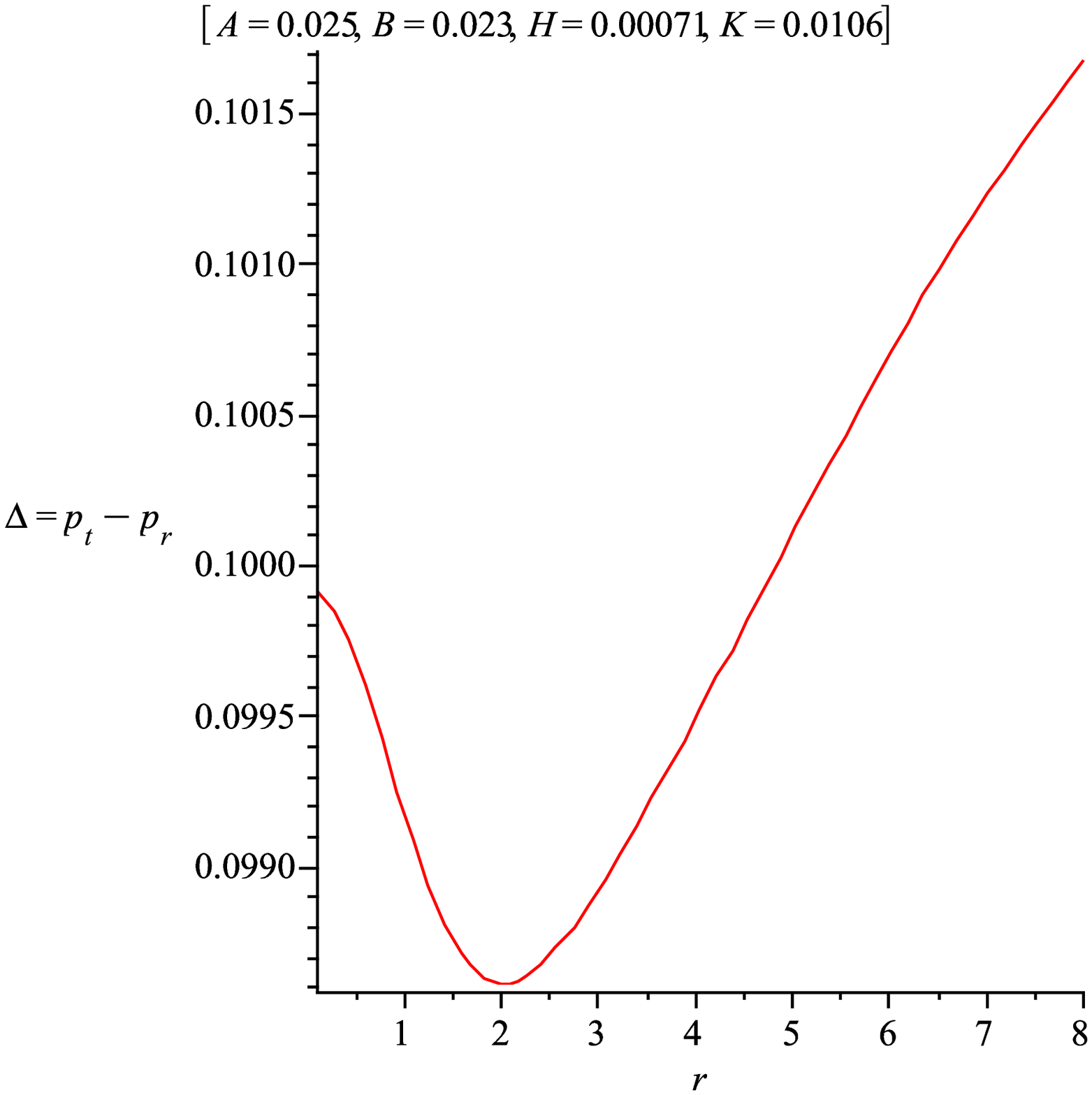}
\caption{The transverse pressure $p_t$ is shown against $r$.}
\label{Fig. 3}
\end{figure*}

\section{TOV equations}

The  generalized Tolman-Oppenheimer-Volkov (TOV) equation as
presented by Ponce de Le\'{o}n \cite{Leon1993} is
\begin{equation}
-\frac{M_G\left(\rho+p_r\right)}{r^2}e^{\frac{\lambda-\nu}{2}}-\frac{dp_r}{dr}+\sigma
\frac{q
}{r^2}e^{\frac{\lambda}{2}}+\frac{2}{r}\left(p_t-p_r\right)=0,
\label{tov}
\end{equation}
where $M_G=M_G(r)$ is the effective gravitational mass inside a
sphere of radius $r$ and $q=q(r)$ is given by (15). The effective
gravitational mass is given by the expression
\begin{equation}
M_G(r)=\frac{1}{2}r^2e^{\frac{\nu-\lambda}{2}}\nu^{\prime},
\label{egm}
\end{equation}
derived from the Tolman-Whittaker formula and the Einstein-Maxwell
equations.

It is important to note that the above equation describes the
equilibrium condition for charged fluid elements subject to
gravitational, hydrostatic and electric forces, plus another force
due to anisotropy factor which is a measure of the pressure
anisotropy of the fluid comprising the charged body. Combined with
(11), (12), (13) and (15), the above equation takes the form
\begin{equation}
 F_g+ F_h+ F_e+ F_a=0,
\end{equation}
where
\begin{equation}
 F_g=-B r\left(\rho+p_r\right),
\end{equation}
\begin{equation}
F_h=-\frac{dp_r}{dr},
\end{equation}
\begin{equation}
F_e=\sigma E e^{\frac{A r^2}{2}},
\end{equation}
\begin{equation}
F_a=\frac{2}{r}\left(p_t-p_r\right).
\end{equation}

The profiles of $F_g$, $F_h$, $F_e$ and $F_a$ for sources are
shown in Fig. 4. This figure indicates that $F_h$ is comparatively
small. Thus the hydrostatic force has a negligible effect in spite
of the static equilibrium is attainable due to pressure
anisotropy, gravitational and electric forces.

\begin{figure*}[htbp]
\includegraphics[scale=.4]{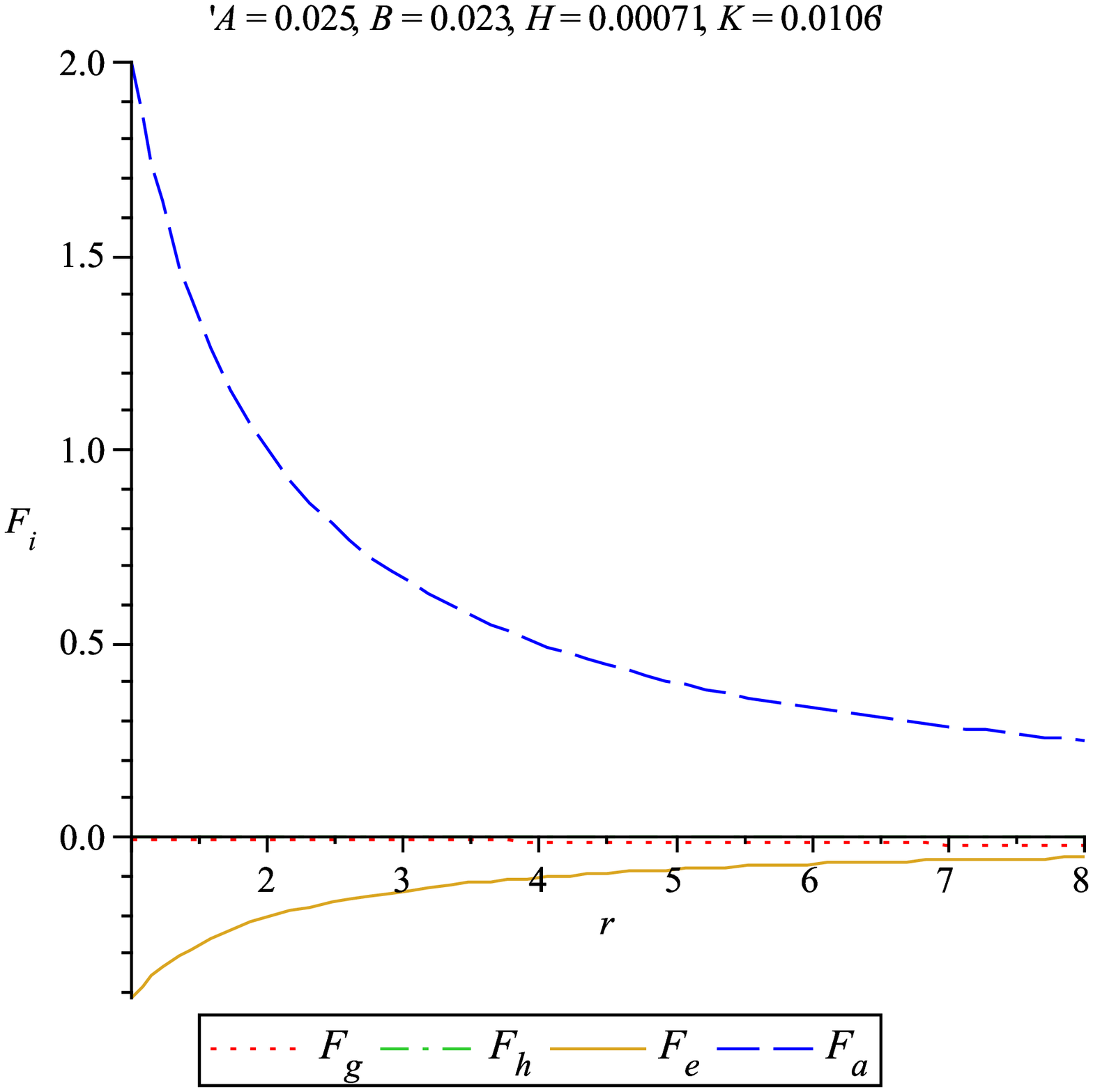}
\caption{Four different forces acting on fluid elements in static
equilibrium is shown against $r$.} \label{Fig. 4}
\end{figure*}

Though several specific equation of states for $p_r(\rho)$ are
used in literature, but a very little is known for the much less
intuitive second equation of state $p_t(\rho)$. The equation of
state parameter $\omega_t \equiv \frac{p_t}{\rho}$ for the
anisotropic object can be obtained directly from equations (11)
and (13), which is given by
\begin{widetext}
\begin{equation}
\frac{p_t}{\rho} \equiv \omega_t = \frac{
\left[2e^{-Ar^2}(B-A)(2+Br^2) -\frac{1}{r^2}\left(1-
e^{-Ar^2}\right) + 8\pi \left(\frac{f +\sqrt{f^2 +
2hK}}{h}\right)\right]}{8\pi \left(\frac{f +\sqrt{f^2 +
2hK}}{h}\right)}
\end{equation}
\end{widetext}

\begin{figure*}[htbp]
\includegraphics[scale=.4]{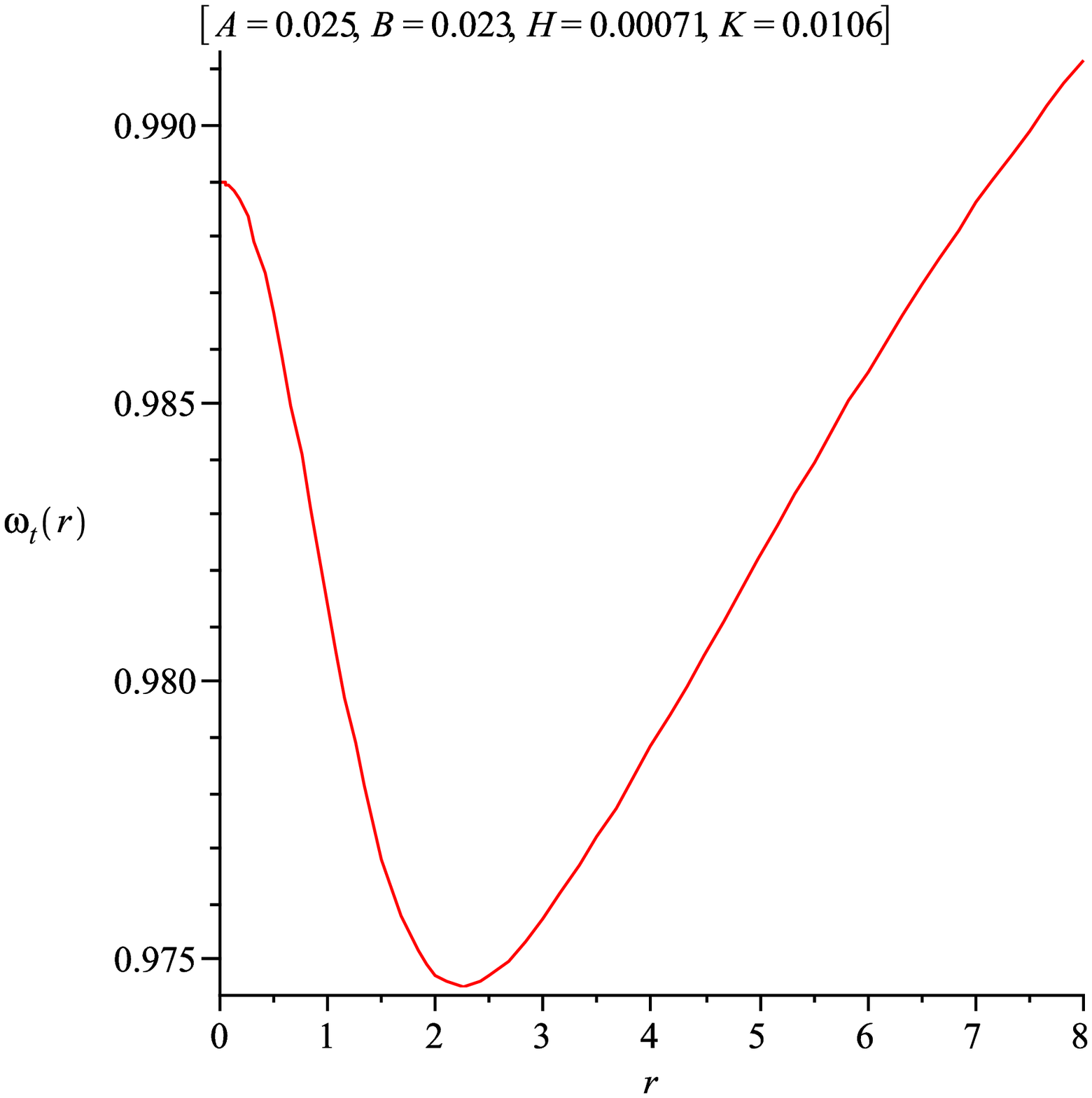}
\caption{The variation of equation of state parameter $\omega_t$
is shown against $r$.} \label{Fig. 5}
\end{figure*}

Fig. 5 shows that the variation of $\omega_t$ against $r$. The
equation of state for   anisotropic charged fluid is positive and
confined within $ 0 \leq \omega_t \leq 1$ i.e. charged fluid is
non exotic in nature.

\section{Energy conditions}

It is well known for the charged fluid that the null energy
condition(NEC), weak energy condition (WEC) and strong energy
condition (SEC) will be satisfied if and only if the following
inequalities hold simultaneously at every point within the source:

\begin{equation}
    \tilde{\rho}+\frac{\tilde{E}^2}{8\pi}\geq 0, \label{ec2}
\end{equation}

\begin{equation}
    \tilde{\rho}+\tilde{p}_r\geq 0, \label{ec1}
\end{equation}

\begin{equation}
    \tilde{\rho}+\tilde{p}_t+\frac{\tilde{E}^2}{4\pi}\geq 0, \label{ec3}
\end{equation}
\begin{equation}
    \tilde{\rho}+\tilde{p}_r+2\tilde{p}_t+\frac{\tilde{E}^2}{4\pi}\geq 0,   \label{ec4}
\end{equation}

Direct plotting of the left sides of (9)-(12) show that these
inequalities are satisfied as well at every $r$ (see Fig. 6).

\begin{figure*}[htbp]
\includegraphics[scale=.4]{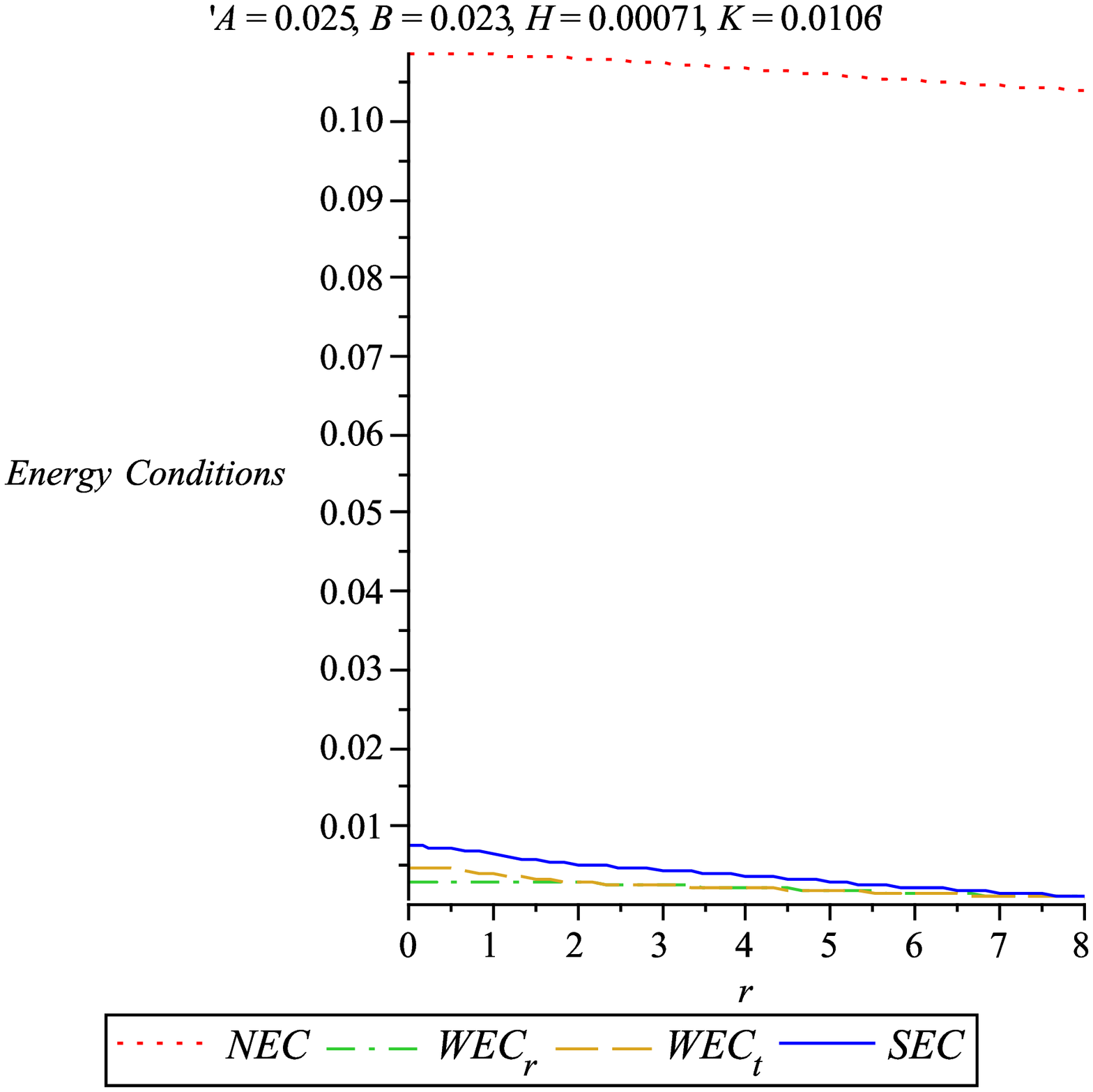}
\caption{The variation of left hand side of the expressions of
energy conditions are shown against $r$.} \label{Fig. 6}
\end{figure*}

At this point we feel it is required to determine whether specific
choices of mass, charge and radius lead to solutions satisfying
the above energy conditions at $r=0$. A close observation of the
equations (\ref{eq:A}) and (\ref{eq:B}) suggest us to adopt here
some adimensional quantities which can be defined as $\alpha=a^2
A$ and $\beta=a^2 B$. We restrict our attention to solutions
satisfying $\alpha \geq 0$, $\beta \geq 0$, $\beta \leq 2\alpha$.
These solutions satisfy the four energy conditions of general
relativity, viz., Null Energy Condition (NEC), Weak Energy
Condition (WEC), Strong Energy Condition (SEC) and Dominant Energy
Condition (DEC).

For a sphere of radius $a$, mass $M$ and charge $Q$, equations
(2.13) - (2.15) in \cite{Junevicus1976} (equations (\ref{eq:A}) -
(\ref{eq:C}) in our present case) are alternatively expressed in
the adimensional forms
\begin{equation}
\alpha=-\ln\left(1-2\mu+\chi^2\right), \label{ea}
\end{equation}
\begin{equation}
\beta=\frac{\mu-\chi^2}{1-2\mu+\chi^2}, \label{eb}
\end{equation}
\begin{equation}
C=\ln\left(1-2\mu+\chi^2\right)-\frac{\mu-\chi^2}{1-2\mu+\chi^2},
\label{ec}
\end{equation}
where $\mu=\frac{M}{a},\chi=\frac{|Q|}{a}$. It is very important
that the field equations can eventually be expressed in terms of
these adimensional constants, the adimensional variables
$\tilde{\rho}=a^2\rho, \tilde{p_r}=a^2p_r, \tilde{p_t}=a^2p_t,
\tilde{\epsilon}=a^2\epsilon$, and the adimensional radial
coordinate $x=\frac{r}{a}$. We have seen that particular values of
the adimensional parameters $\mu, \chi$ determine the adimensional
KB constants $\alpha$, $\beta$, $C$ which in turn determine
$\lambda$ and $\nu$ at every $x\in [0,1]$. The values of $\mu$ and
$\chi$ are restricted by the condition that no horizon is included
in the external region described by the RN metric.

We consider all possible roots of the equation $g_{00}=0$. The
radius of the charged sphere $a$ is big enough so that no horizons
are included in the external RN metric. The three possible cases
are:

\subsection{Two real roots} $\mu^2>\chi^2$.\\

We choose $1>\mu+\sqrt{\mu^2-\chi^2}$. Therefore, $\chi<1$ and
$\mu$ satisfies $\chi<\mu<\frac{1+\chi^2}{2}$.

\subsection{One real root} $\mu=\chi$.\\

We choose $1>\mu=\chi$.

\subsection{No real roots} $\mu<\chi$, otherwise arbitrary.\\

The selected values of $\mu$ and $\chi$ determine values of
$\alpha$ and $\beta$ which should satisfy the energy conditions.
Another acceptability condition is that $\tilde{\epsilon}(x)\geq
0$ for every $x\in[0,1]$.

The arising expressions for $\rho$ and $p_r$ and $p_{t}$ can be
evaluated at $r=0$. Hence we find the energy density and pressures
at $r=0$ as
\begin{equation}
\rho_0=\frac{3A}{8\pi}=\frac{3\alpha}{8\pi a^2}, \label{eq:rho02}
\end{equation}
\begin{equation}
p_{r0}=\frac{1}{2}p_{t0}=\frac{2B-A}{8\pi}=\frac{2\beta-\alpha}{8\pi
a^2}. \label{eq:pr0}
\end{equation}

Now, The energy conditions \cite{Hawking1973} at the centre can
be written as :\\

(i) NEC: $p_{r0} + \rho_0 \geq 0$ $\Rightarrow$ $ \alpha + \beta
\geq 0 $ \\

(ii) WEC: $p_{r0} + \rho_0 \geq 0$ $\Rightarrow$ $ \alpha + \beta
\geq 0 $ \& $\rho_0 \geq 0$ $\Rightarrow$ $ \alpha \geq 0 $\\

(iii) SEC: $p_{r0} + \rho_0 \geq 0$ $\Rightarrow$ $ \alpha + \beta
\geq 0 $ \& $3p_{r0} + \rho_0 \geq 0$ $\Rightarrow$ $ \beta \geq 0
$\\

(iv) DEC: $ \rho_0 > \mid p_{r0} \mid $ $\Rightarrow$ $ 2\alpha
\geq \beta \geq \alpha$\\

The characterization of dark energy fluid is the violation of one
of the SEC, more specifically, that one related the Raychaudhuri
equation \cite{Chan2008a,Chan2008b}. If the second of the WEC is
violated, we have a phantom dark energy fluid.

Now, EOS at $r=0$ is
\begin{equation}
\frac{p_{r0}}{\rho_0} = m
\end{equation}
where,
\begin{equation}
m = \frac {2\beta - \alpha}{3\alpha} \leq \frac{\beta}{2\alpha}
\leq 1
\end{equation}

Notice that $\rho$ and $p_r$ are decreasing functions of $r$
(these can be shown by plotting the graphs of $\rho$ and $p_r$ or
one can find $\frac{d\rho}{dr} < 0$ and $\frac{dp_r}{dr} < 0$ i.e.
$\rho$ and $p$ are decreasing functions of $r$). Since, at $r =
0$, they assume fixed values, $\rho_0$ and $p_{r0}$. So, $\rho$
and $p$ has a maximum at $r=0$. We have checked that
$\rho_0^\prime = 0 $, $ p_{r0}^\prime = 0 $ and $\rho_0^{\prime
\prime } < 0$, $p_{r0}^{\prime \prime} < 0$.

\section{Stability}

Bertolami and P{\'a}ramos \cite{Bertolami2005} argue that if one
assumes that the GCG tends to a smooth distribution over space
then most density perturbations tend to be flattened within a time
scale related to their initial size and the characteristic speed
of sound.

One of the important ``physical acceptability conditions'' for
anisotropic matter are the squares of radial and tangential sound
speeds, defined by
\begin{equation}
v_{sr}^2=\frac{dp_r}{d\rho} = H + \frac{K}{\left[\frac{f
+\sqrt{f^2 + 2hK}}{h}\right]^2}. \label{eq:velo}
\end{equation}

and

\begin{widetext}
\begin{equation}
v_{st}^2=\frac{dp_t}{d\rho} = 1- \frac{\sqrt{f^2+2hK}}{4\rho r
A(A+B)}
 \left[ \frac{2}{r^2} e^{Ar^2} - 4
rA(B-A)(1+Br^2)-\frac{2}{r^3} -\frac{2A}{r}\right]
 \end{equation}
\end{widetext}

should be less than the speed of light
\cite{Herrera1992,Abreu2007}.

From the above equation (\ref{eq:velo}), an important aspect can
be observed that the squared of  radial sound velocity is always
positive  irrespective of matter density and hence this is always
positive even in the case of exotic matter. The Figs 7 and 8 show
that these parameters satisfy the inequalities $0\leq v_{sr}^2
\leq 1$ and $0\leq v_{st}^2 \leq 1$ everywhere within the charged
fluid.

\begin{figure*}[htbp]
\includegraphics[scale=.4]{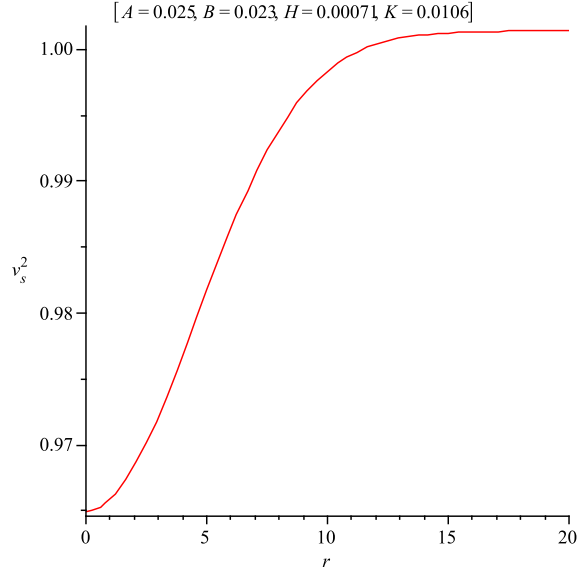}
\caption{The variation of radial sound speed $v_{sr}^2 $ is shown
against $r$.} \label{Fig. 7}
\end{figure*}

\begin{figure*}[htbp]
\includegraphics[scale=.4]{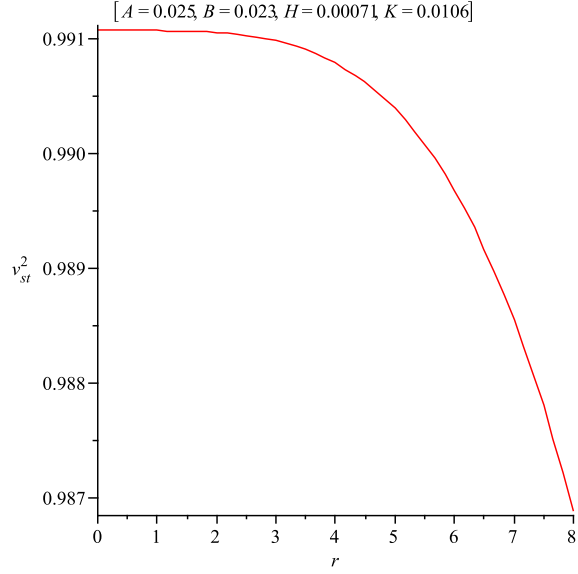}
\caption{The variation of tangential sound speed $v_{st }^2 $ is
shown against $r$.} \label{Fig. 8}
\end{figure*}

Now, we use Herrera's  approach \cite{Herrera1992} to identify
potentially unstable or stable  anisotropic matter configuration
known as the concept of cracking (or overturning). Since, $0\leq
v_{sr}^2 \leq 1$ and $0\leq v_{st}^2 \leq 1$,  therefore,
according to \cite{Herrera1992,Andreasson1992} , $\mid v_{st}^2 -
v_{sr}^2 \mid \leq 1 $. The Fig. 9 of the model also supports
this.

\begin{figure*}[htbp]
\includegraphics[scale=.4]{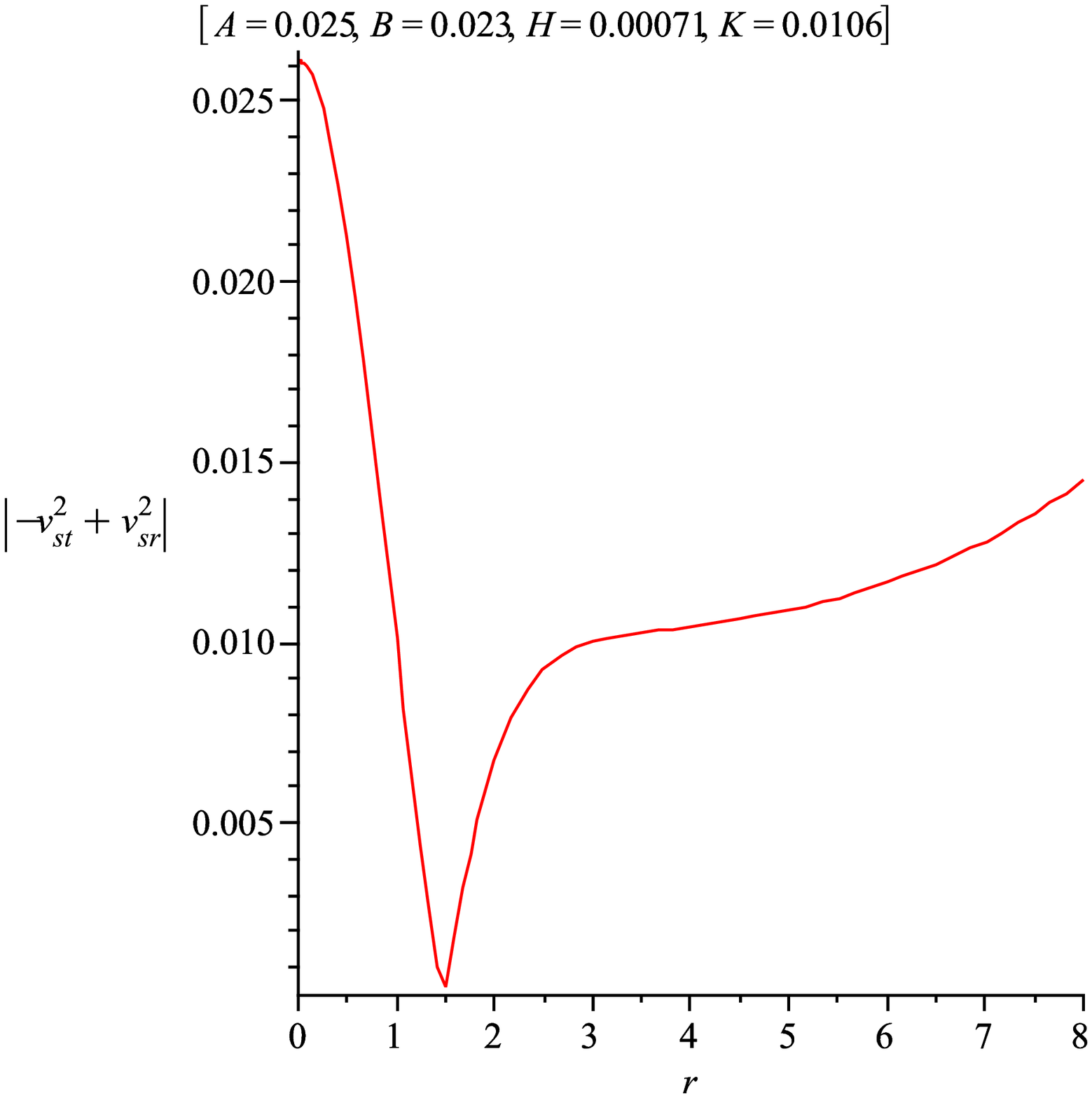}
\caption{The variation of $\mid v_{st}^2 - v_{sr}^2 \mid $ is
shown against $r$.} \label{Fig. 9}
\end{figure*}

Now,\[  -1 \leq  v_{st}^2 - v_{sr}^2  \leq  1\] implies

\[ -1 \leq  v_{st}^2 -
v_{sr}^2  \leq 0 ~ , ~~~~~~~~~~~~~~~  potentially ~~ stable\]
\[ 0 <   v_{st}^2 -
v_{sr}^2  \leq 1 ~ ,~~~~~~~~~~~~~~~  potentially ~~ unstable\]

One can note that the region for which $v_{st}^2 <  v_{sr}^2 $ is
potentially stable region and the region for which $v_{st}^2 >
v_{sr}^2 $ is potentially unstable region. If $v_{st}^2 - v_{sr}^2
$ keeps the same sign everywhere within a matter distribution, no
cracking will occur. The curve profile (Fig. 10) for $v_{st}^2 -
v_{sr}^2 $ indicates that there is a change of sign and thus
alternating potentially unstable to stable region within the
distribution.

\begin{figure*}[htbp]
\includegraphics[scale=.4]{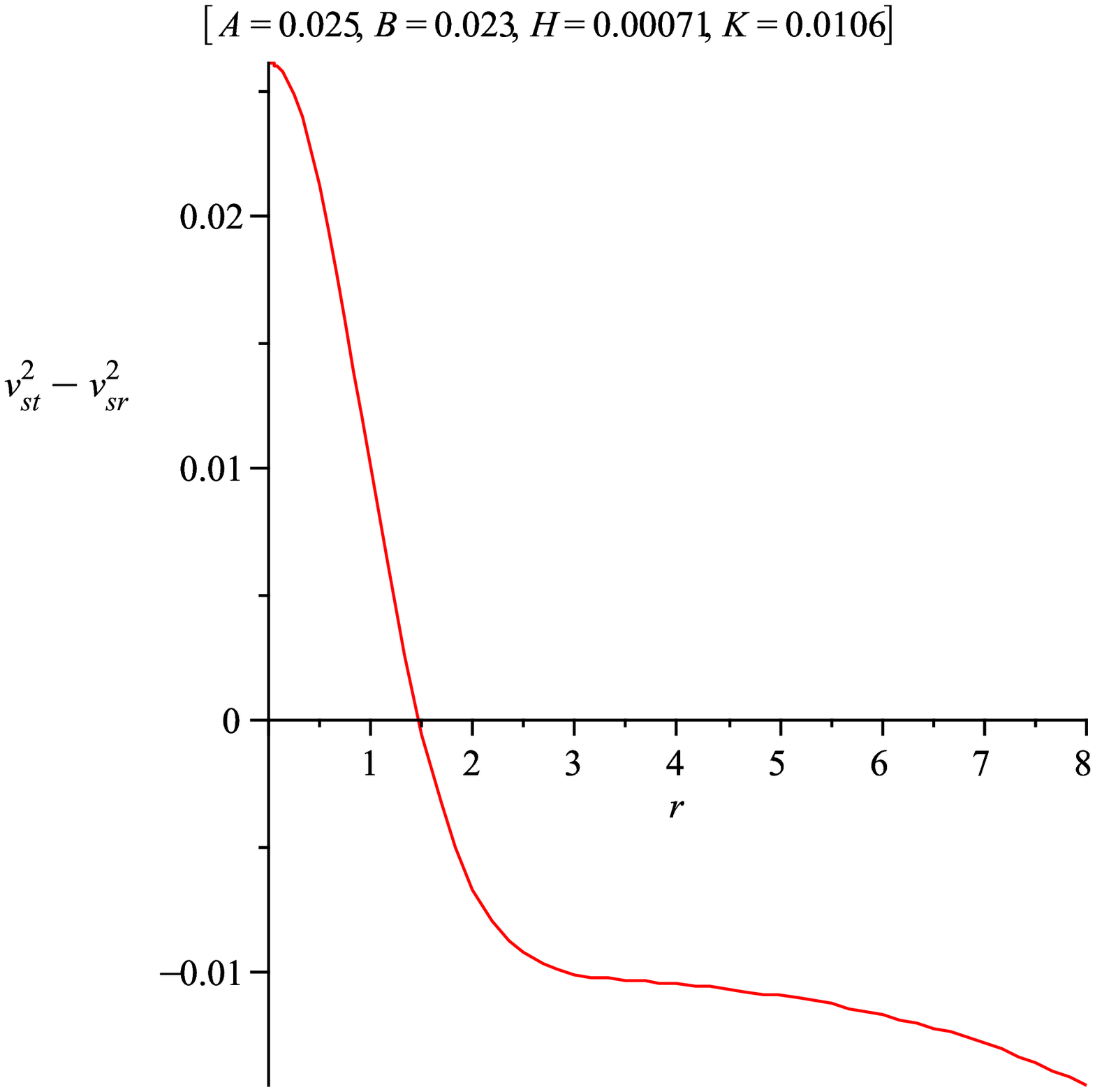}
\caption{The variation of $ v_{st}^2 - v_{sr}^2 $ is shown against
$r$.} \label{Fig. 10}
\end{figure*}

\section{Minimum mass-radius relation}

From, the above analysis indicates that our model is very much
unstable within radius $1.5$ unit. But, the the configuration is
stable within $1.5< r \leq 8$. In a recent paper, Andr\'{e}asson
\cite{Andreasson1992} has discovered a surprising result as:

\[ \sqrt{M} < \frac{\sqrt{R}}{3} + \sqrt{\frac{ R }{9} + \frac{Q^2}{3R}}\]

for a lower bound on the radius R of a charged sphere with mass
$M$ and charge $Q$.

The inequality is shown to hold for any solution which satisfies
$p_r+ 2p_t  \leq \rho $.

The plot (Fig. 11) for $p_r+ 2p_t  - \rho $ against $r$ indicates
that in the region $1.5< r \leq 8$, $p_r+ 2p_t  - \rho $ is
negative. Since, our model is stable within $1.5 < r \leq 8$, so
Andr\'{e}asson's relation holds good for our model.

\begin{figure*}[htbp]
\includegraphics[scale=.4]{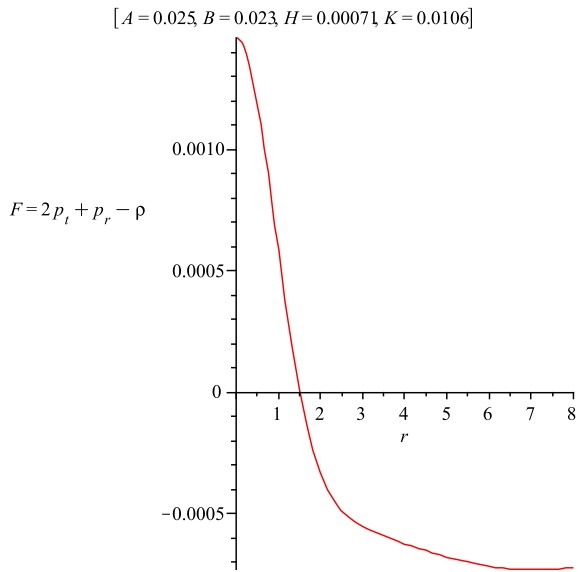}
\caption{The variation of $2p_t+p_r - \rho$ is shown against $r$.}
\label{Fig. 11}
\end{figure*}

It would be interesting to made some comments regarding opposite
situations with the maximum mass-radius relation. By using the
static spherically symmetric gravitational field equations
Buchdahl \cite{Buchdhal1959} has obtained an absolute constraint
of the maximally allowable mass-radius for isotropic fluid spheres
of the form $\frac{2M}{R} < \frac{8}{9}$ (for a generalized
expression we refer the work of Mak et al. \cite{Mak2001}).

It is worthwhile to calculate effective gravitational mass which
due to the contribution of the energy density $\rho$ of the matter
and the electric energy density $\frac{E^2}{8\pi}$ and can be
expressed as
\begin{widetext}
\begin{equation}
M_{effective} = 4 \pi \int_0^R \left[ \rho
+\frac{E^2}{8\pi}\right] r^2 dr = \frac{1}{2}R +
\frac{1}{\sqrt{A}} \gamma \left( \frac{3}{2} , AR^2\right)-
\frac{1}{2\sqrt{A}} \gamma \left( \frac{1}{2} , AR^2\right)
\end{equation}
\end{widetext}

where $\gamma \left( \frac{3}{2}, AR^2\right)$ is the lower
incomplete gamma function. In Fig. 12, we plot the mass-radius
relation. We also plot $\frac{M _{effective}}{R}$ against $R$ (see
Fig. 13) which shows that the ratio $\frac{M _{effective}}{R}$ is
decreasing even if the radius is increasing with the mass.

\begin{figure*}[htbp]
\includegraphics[scale=.4]{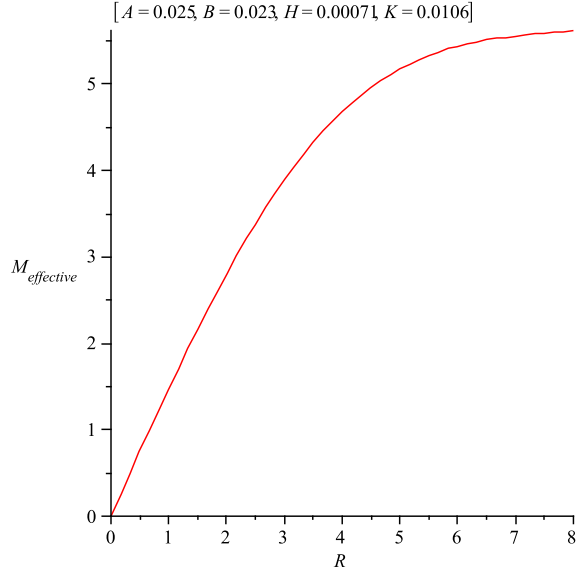}
\caption{The variation of $M_{effective}$  is shown against R.}
\label{Fig. 12}
\end{figure*}

\begin{figure*}[htbp]
\includegraphics[scale=.4]{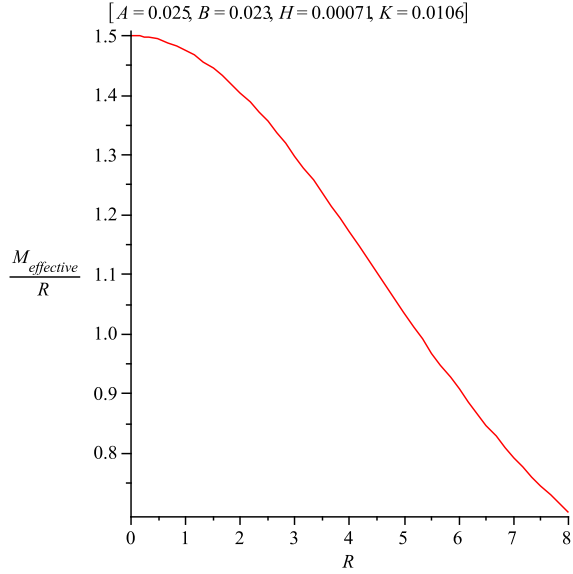}
\caption{The variation of $\frac{M _{effective}}{R}$  is shown
against R.} \label{Fig. 13}
\end{figure*}

According to Ponce de Le\'{o}n \cite{Leon1993} the energy
conditions require
\[ e^{-\lambda} \leq 1 ~~, ~~ ~~ \nu^\prime \geq 0 \]. These relations
lead to a maximum charge is as follows:

\[ R \geq \frac{q^2}{M}  \]

where $M$ and $q$ represent the total mass and charge of the
charged sphere of radius $R$.

Assuming $R = 8$ Km and $\frac{M}{R} = 0.5$, we find from equation
(15) as $q^2 = 0.20372$. Thus our model satisfies the Ponce de
Le\'{o}n's condition \cite{Leon1993}.

In this connection we add that the mass-radius-charge relation for
compact astrophysical objects plays an important role in many
physical processes. The strong gravitational field due to the
density of the matter inside the stars indicates that a strong
electric field due to the electric charge is possible to exist.
The effect of electric charge in compact stars assuming that the
charge distribution is proportional to the mass density were
studied recently by several authors
\cite{Ray2003,Ghezzi2005,Ray2007a,Bohmer2007}.

\section{Conclusions}
In this paper we are checking the energy conditions only at the
center of the charged sphere. It would be convenient to extend the
analysis to other points within the sphere. A series method like
the one used in equations (32) and (34) by KB \cite{Krori1975}
might be useful.

Unlike the work of Bertolami and P{\'a}ramos \cite{Bertolami2005}
where they have used the generalized Chaplygin gas (GCG) EOS in
special reference to anisotropic pressure we generate the
solutions for KB metric under Einstein-Maxwell space-time. So a
natural question - Does the result go over into the solution of
Bertolami et al. for isotropic stresses? The straight forward
answer is No. This is because we have extended KB approach
assuming singularity free form of the metric ansatz to charged
anisotropic source with non linear, Chaplygin type equation of
state. Therefore, whether our solution corresponds to Chaplygin
dark star needs a special verification, specifically whether
charged Chaplygin dark star does exist demands further
investigation. In a similar fashion one may raise the question
that does the result go over into an exact solution for an EOS $p=
H\rho~ (K=0$)? The answer this time is affirmative, as our
solution coincides with the solution obtained by Varela et al.
\cite{Varela2010} for an EOS $p= H\rho~ (K=0$). One can see easily
that our results go over into the expressions obtained by Varela
et al. with $\alpha_1 =0$. Also, It may be interesting to
extrapolate the present investigation to the astrophysical bodies,
specially quark or strange stars with radius around $8$ km. \\

\section*{APPENDIX: ANALYSIS OF JUNCTION CONDITIONS}
\def\theequation{A.\arabic{equation}}

We can note that the metric coefficients continuous at the
junction i.e. at $S$ where $r=a$. However, this does not mean that
the metric coefficients be differentiable at the junction. The
affine connection may be discontinuous there. The above  statement
can be quantified in terms of second fundamental form of the
boundary.

The second fundamental forms associated with the two sides of the
shell are \cite{Israel1966,Rahaman2007a,Rahaman2007b,Rahaman2007c}
\begin{equation}K_{ij}^\pm =  - n_\nu^\pm\ \left[ \frac{\partial^2X_\nu}
{\partial \xi^i\partial \xi^j } +
 \Gamma_{\alpha\beta}^\nu \frac{\partial X^\alpha}{\partial \xi^i}
 \frac{\partial X^\beta}{\partial \xi^j }\right] |_S
 \end{equation}
where $ n_\nu^\pm\ $ are the unit normals to $S$ and can be given
by
\begin{equation} n_\nu^\pm =  \pm   \left| g^{\alpha\beta}\frac{\partial f}{\partial X^\alpha}
 \frac{\partial f}{\partial X^\beta} \right|^{-\frac{1}{2}} \frac{\partial f}{\partial X^\nu}
 \end{equation}
with $ n^\mu n_\mu = 1$. Here, $\xi^i$ are the intrinsic
coordinates on the shell with $f =0$ is the parametric equation of
the shell $S$ and $-$ and $+$ corresponds to interior (our) and
exterior (RN). It is to be noted that since the shell is
infinitesimally thin in the radial direction there is no radial
pressure. Using Lanczos equations
\cite{Israel1966,Rahaman2007a,Rahaman2007b,Rahaman2007c}, one can
find the surface energy term $\Sigma$ and surface tangential
pressures $ p_\theta = p_\phi \equiv p_t$ as
\begin{equation}
\Sigma =  - \frac{1}{4\pi a}\left[ \sqrt{e^{-\lambda}}\right]_-^+,
\end{equation}
\begin{equation}
p_t =   \frac{1}{8\pi a}\left[ \left( 1 + \frac{a \nu^\prime
}{2}\right) \sqrt{e^{-\lambda}}\right]_-^+.
\end{equation}

The metric functions are continuous on $S$, then one finds
\begin{equation}
\Sigma = 0,
\end{equation}
\begin{widetext}
\begin{eqnarray}
p_t = \frac{1}{8\pi a} \left[ \left(\frac{M}{ a} - \frac{Q^2}{
a^2}\right) {\sqrt{1 - \frac{2M}{a}+ \frac{Q^2}{ a^2}}}
-Aa^2{\sqrt{1 - \frac{2M}{a}+ \frac{Q^2}{ a^2}}}\right].
\end{eqnarray}
\end{widetext}

Hence one can match our interior solution with an exterior RN
solution in the presence of a thin shell. The whole space-time is
given by our metric and RN metric which are joined smoothly.\\

\section*{Acknowledgments} FR and SR are thankful to the authority of
Inter-University Centre for Astronomy and Astrophysics, Pune,
India for providing them Visiting Associateship under which a part
of this work was carried out. FR is also thankful to PURSE for
providing financial support. We are very grateful   to an
anonymous referee for his/her insightful comments that have led
to significant improvements, particularly on the interpretational
aspects.

\end{document}